\documentclass[aps,prx,10pt,
superscriptaddress,
twocolumn,
longbibliography,
showpacs]{revtex4-2}

\usepackage[utf8]{inputenc}
\usepackage[english]{babel}  

\usepackage{amssymb,amsmath,amsfonts,amsthm}
\usepackage{mathtools}
\usepackage{verbatim}
\usepackage{enumerate}
\usepackage{graphicx}
\graphicspath{{pics/}}
\usepackage{hyperref}
\usepackage{bbm}
\usepackage{booktabs}

\usepackage[caption=false]{subfig}

\usepackage{color}
\definecolor{dred}{rgb}{.8,0.2,.2}
\definecolor{ddred}{rgb}{.8,0.5,.5}
\definecolor{dblue}{rgb}{.2,0.2,.8}
\definecolor{dgreen}{rgb}{.2,0.5,.2}

\theoremstyle{plain}

\theoremstyle{definition}

\newcommand{\be}{\begin{equation}}
\newcommand{\ee}{\end{equation}}

\usepackage{tikz}
\usepackage{braids}
\usepackage{xifthen}

\usepackage{euscript}

\newcommand\Zb            {\mathbb{Z}}

\newcommand\CD           {\EuScript{D}}

\usepackage{array}
\newcolumntype{C}[1]{>{\centering\let\newline\\\arraybackslash\hspace{0pt}}m{#1}}

\begin{document}
	
\title{Measuring the Unique Identifiers of Topological Order Based on Boundary-Bulk Duality and Anyon Condensation}

    
  \author{Yong-Ju Hai}
    \thanks{These authors contributed equally to this work.}
    \affiliation{Department of Physics and Institute for Quantum Science and Engineering, Southern University of Science and Technology, Shenzhen 518055, China}

  \author{Ze Zhang}
    \thanks{These authors contributed equally to this work.}
    \affiliation{Department of Physics and Institute for Quantum Science and Engineering, Southern University of Science and Technology, Shenzhen 518055, China}
    
  \author{Hao Zheng}       
     \affiliation{Department of Physics and Institute for Quantum Science and Engineering, Southern University of Science and Technology, Shenzhen 518055, China}
     \affiliation{Guangdong Provincial Key Laboratory of Quantum Science and Engineering, Shenzhen 518055, China}
  
  \author{Liang Kong}
     \email{kongl@sustech.edu.cn }  
     \affiliation{Department of Physics and Institute for Quantum Science and Engineering, Southern University of Science and Technology, Shenzhen 518055, China}   
      \affiliation{Guangdong Provincial Key Laboratory of Quantum Science and Engineering, Shenzhen 518055, China}

  \author{Jiansheng Wu}
     \email{wujs@sustech.edu.cn}
     \affiliation{Department of Physics and Institute for Quantum Science and Engineering, Southern University of Science and Technology, Shenzhen 518055, China}  
     \affiliation{Guangdong Provincial Key Laboratory of Quantum Science and Engineering, Shenzhen 518055, China}      

  \author{Dapeng Yu}   
     \affiliation{Department of Physics and Institute for Quantum Science and Engineering, Southern University of Science and Technology, Shenzhen 518055, China}
     \affiliation{Guangdong Provincial Key Laboratory of Quantum Science and Engineering, Shenzhen 518055, China}    


%
	

\begin{abstract}

A topological order is a new quantum phase that is beyond Landau's symmetry-breaking paradigm. Its defining features include robust degenerate ground states, long-range entanglement and anyons. It was known that $R$- and $F$-matrices, which characterize the fusion-braiding properties of anyons, can be used to uniquely identify topological order. In this article, we explore an essential question: how can the $R$- and $F$-matrices be experimentally measured? By using quantum simulations based on a toric code model with boundaries and state-of-the-art technology, we show that the braidings, i.e. the $R$-matrices, can be completely determined by the half braidings of boundary excitations due to the boundary-bulk duality and the anyon condensation. The $F$-matrices can also be measured in a scattering quantum circuit involving the fusion of three anyons in two different orders. Thus we provide an experimental protocol for measuring the unique identifiers of topological order. 
Our experiments are accomplished by means of our NMR quantum computer at room temperature. We simplify the toric code model in 3, 4-qubit system and our measured $R$- and $F$-matrices are all consistent with the theoretical prediction.

\end{abstract}

\maketitle

\section{Introduction}

Topological orders are defined for gapped many-body systems at zero temperature. They were first discovered in  2-dimensional (2d) fractional quantum Hall systems, and are new types of quantum phases beyond Landau's symmetry-breaking paradigm~\cite{XGW, WenNiu, XGW2, MS89, Kitaev06, Preskill, Levin, Tsui, Laughlin, Wu, Wu2, Wu3, Girvin, Zhang, Blok, Read, WenZee}. Not only they challenge us to find a radically new understanding of  phases and phase transitions, but also  provide the physical foundation of fault-tolerant quantum computers~\cite{Kitaev,Dennis, Freedman,Chetan,bk}.

\begin{figure*}
	\centering
	\includegraphics[scale=0.95]{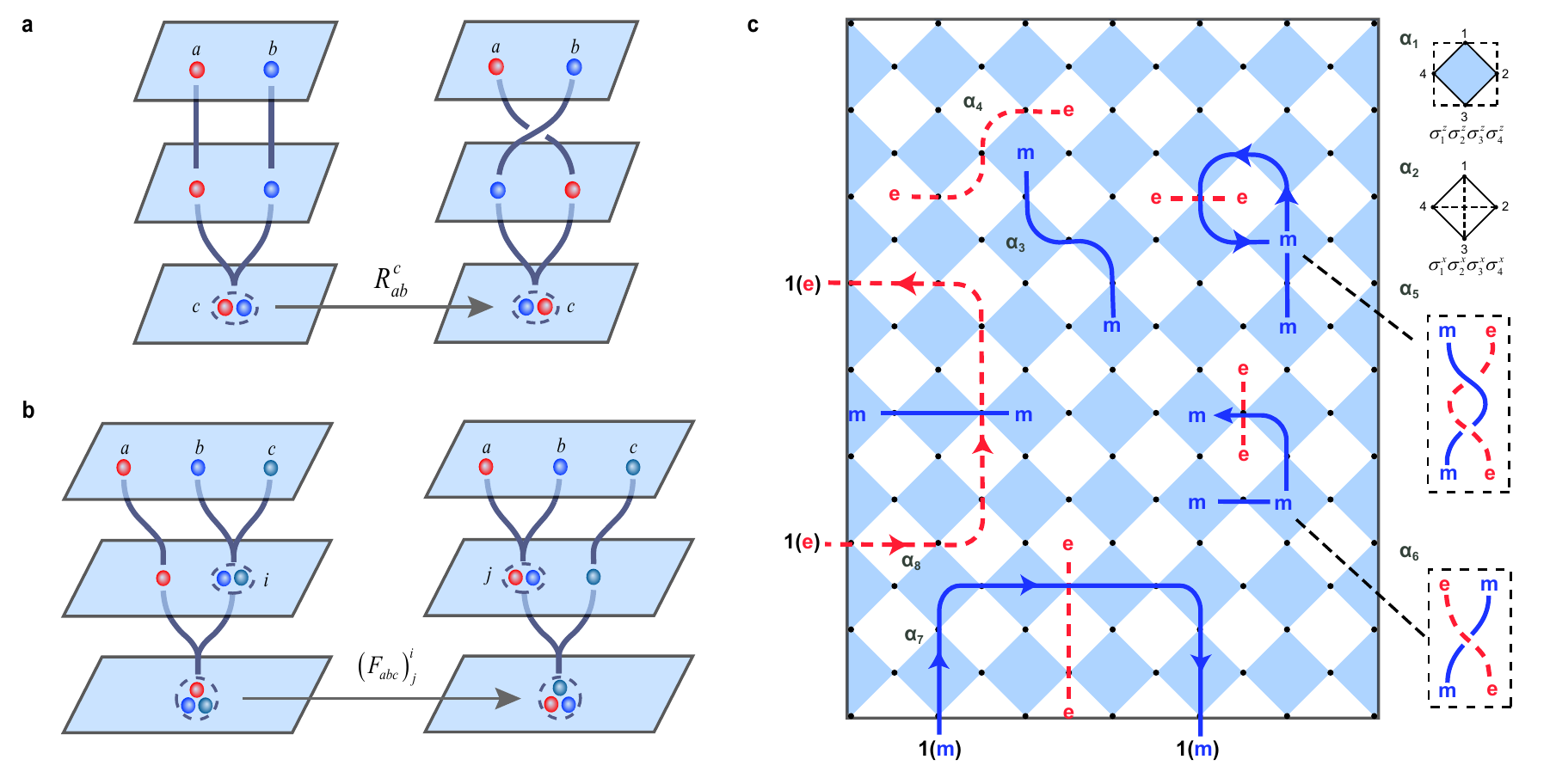}
	\caption{{\bf Graphical representation of the $R$-matrices, the $F$-matrices and  a toric code lattice with gapped boundaries.} 
		{\bf a}, Braiding of two anyons and the definition of the $R$-matrices;
		{\bf b}, Fusion of three anyons in different orders and the definition of the $F$-matrices;
		{\bf c}, Toric code lattice with boundaries.  $\alpha_1$ and $\alpha_2$ are the two elementary plaquettes, the blue plaquette and the white plaquette. $\alpha_3$ and $\alpha_4$ are the string operators for generating an $m$ anyon pair and an $e$ anyon pair, respectively.  $\alpha_5$ and $\alpha_6$ visualize the double braiding and braiding operations. Double braiding corresponds to moving an $m$ ($e$) anyon around an $e$ ($m$) anyon along a full circle, which generates an overall phase of $-1$. 
		Braiding corresponds to the  exchange of an $m$ anyon and an $e$ anyon: If they are in the bulk, the state after braiding is different from the initial state.  $\alpha_7$ and $\alpha_8$ are half braidings at the white (smooth) and blue (rough) boundaries. 
		At a white boundary, an $m$ anyon is equivalent to vacuum state $1$.  Dragging an $m$ anyon from the vacuum state, moving it around a boundary excitation $e$  along a semicircle and pushing it back out results in  a phase difference of $-1$.  The case is similar for a blue boundary.}
	\label{fig1}
\end{figure*}

 The first fundamental question is how to characterize and measure a topological order precisely. Important progress has been made toward achieving this goal, such as the measurement of modular data, such as $S$-matrices, $T$-matrices and topological entanglement entropy~\cite{RSW, Duan, Wang, Du, Lu, Preskill,Levin, TEE}. This motivated a folklore belief among experts that the modular data might be complete~\cite{Zoo}. A recent mathematical result~\cite{MS}, however, suggests that it is incomplete which means that different topological order might have the same modular data. The complete characterization of topological order should be $R$- and $F$-matrices.
 A 2d topological order permits particle-like topological excitations, called anyons. When two anyons are braided (exchanged), a ``phase factor'', (or a matrix) called an $R$-matrix, is presented in their wavefunctions (as illustrated in Fig.~\ref{fig1}a). Furthermore, two anyons can be fused together to produce a new anyon, from one or several possible outcomes.  If we fuse three anyons $a$, $b$ and $c$ in two different ways, $((ab)c)$ and $(a(bc))$, where parentheses indicates the order of the fusion, the first set of fused states spans the same Hilbert space as the second one. The $F$-matrix is the transformation matrix between these two bases (as illustrated in Fig.~\ref{fig1}b). Mathematically, it was proved that  the $R$-matrices and $F$-matrices uniquely determine the topological order~\cite{MS89,Kitaev06}. Consequently, they can serve as the unique identifiers of a topological order.

Therefore, the essential question is whether $R$-matrices (braidings) and $F$-matrices are physically measurable? The difficulty of measuring the $R$-matrices are twofold. The first one is that different (gauge choices of) $R$-matrices might define the same topological order. So they are not unique and seem not to be measurable quantities. The second one is that by the definition of braiding if we move one anyon along a semicircle around another anyon of a different kind in the bulk, the spatial configuration of the final state is so different from the initial one that there's no well-defined phase factor. 

In this paper, we address this fundamental question and overcome these two difficulties using the boundary-bulk duality and the anyon condensation on the boundaries~\cite{KK,KWZ,KL,kz19a,kz19b},
which indicates that the braidings among bulk anyons are determined by the half braidings among boundary excitations, and these half braidings can be measured by moving one boundary excitation around another one along a semicircle near the boundary. We find different (but gauge equivalent) $R$-matrices are associated to different boundaries of the same 2d topological order.
When the boundary is gapped, certain bulk anyons are condensed on the boundary, thus can be created or annihilated on the boundary by local operators.  
By creating an anyon at the boundary then half braiding it with another anyon then annihilating it on the boundary, 
we obtain a final state which differs from the initial state only by a phase factor ($R$-matrix), thus overcome the second difficulty. We also show that the $F$-matrices are measurable using a  quantum circuit involving the fusion of three anyons in different orders. 
As a consequence, we provide a protocol for experimentally measuring the $R$-matrices and $F$-matrices.

\section{Measurement of $R$-matrices and $F$-matrices of topological order}


\subsection{Unique identifier of topological order}

As mentioned above, topological orders can be characterized by their particle-like excitation, anyons.
Anyon models describe the behavior of the topological sectors in a gapped system whose phase is subject to (2+1)d Topological Quantum Field Theory (TQFT). They can be described by unitary modular tensor categories (UMTC) mathematically~\cite{Kitaev06}. We label the anyons in a finite set $\mathcal{C}$ by $a,b,c,\dots$, and the vacuum sector by $\bf{1}$. 

There are several major concepts in anyon models. The first is fusion between two anyons, where two anyons are combined together, or fused, to give an anyon. It is formulated by a fusion rule:
\begin{equation}
a \otimes b = \sum_c N_{ab}^c c,
\end{equation}
the integer multiplicity $N_{ab}^c$ gives the number of times $c$ appears in the fusion outcomes of $a$ and $b$.  
The quantum dimension $d_a$ of an anyon $a$ is defined through the fusion rule $d_a d_b=\sum_c N_{ab}^c d_c$. It represents the effective number of degrees of freedom of the anyon. For a non-Abelian anyon $a$, we have some $b$ such that $\sum_c N_{ab}^c > 1 $ and $d_a > 1$. If $\sum_c N_{ab}^c=1$ for every $b$, then the anyon $a$ is Abelian and we have $d_a=1$. 

The integer $N_{ab}^c$ is the number of different ways in which $a$ and $b$ fuse to $c$ hence is the dimension of the Hilbert space (the fusion space) $\hom(a\otimes b,c)$. We choose an orthonormal basis of $\hom(a\otimes b,c)$, denoted diagrammatically by 
$  \{ \begin{tikzpicture}[scale=0.5,font=\footnotesize,anchor=mid,baseline={([yshift=-.5ex]current bounding box.center)}]
 \def\height{1.2}
 \node at (0.5, 1.4) {a};
 \node at (1.5, 1.4) {b};
 \draw (0.5, 1.2) to (1, 0.7);
 \draw (1.5, 1.2) to (1, 0.7);
 \draw (1, 0.7) to (1, 0);
 \node at (1.3, 0) {c};
 \node at (1.3, 0.6) {v};
 \end{tikzpicture} \}_{v=1}^{N_{a b}^{c}} $.

Another important concept is braiding, which corresponds to the exchange of two anyons. The braiding operation of $a$ and $b$ in the fusion channel $c$ is represented by
\begin{equation}
\tikzset{every picture/.style={line width=0.75pt}}
\tikzset{every picture/.style={line width=0.75pt}}      
\begin{tikzpicture}[x=0.75pt,y=0.75pt,yscale=-1,xscale=1]
\draw    (200.25,71) -- (215,91) ;
\draw    (215,91) -- (215,116) ;
\draw    (226.98,71) -- (215,91) ;
\draw    (92,70.52) -- (102.35,80.86) ;
\draw    (105.56,82.29) .. controls (99.57,87.17) and (96.17,92.8) .. (104.83,100.31) ;
\draw    (106.59,86.03) .. controls (109.42,88.72) and (112.32,92.28) .. (104.83,100.31) ;
\draw    (104.83,100.31) -- (104.83,118.29) ;
\draw    (117.01,70.84) -- (105.56,82.29) ;
\draw (93,62) node  [align=left] {{\footnotesize a}};
\draw (117,63) node [scale=0.8] [align=left] {b};
\draw (112,118) node [scale=0.8] [align=left] {c};
\draw (201,63) node  [align=left] {{\footnotesize a}};
\draw (228,64) node [scale=0.8] [align=left] {b};
\draw (222,116) node [scale=0.8] [align=left] {c};
\draw (99,101) node [scale=0.8] [align=left] {v};
\draw (208,92) node [scale=0.8] [align=left] {u};
\draw (228,92) node  [align=left] {,};
\draw (153,89) node  [align=left] {};
\draw (153,90) node   {$$};
\draw (155,89) node  [align=left] {$\displaystyle =\sum ^{N^{c}_{ab}}_{u}\left( R^{c}_{ab}\right)^{u}_{v}$};
\end{tikzpicture}
\end{equation}
where $R^c_{a b}$ is the so called $R$-matrix. If $a$ or $b$ is an Abelian anyon so that there exists a unique fusion channel $c=a\otimes b$, the matrix $R^c_{a b}$ is reduced to a number $R_{a b}$. In particular, in an Abelian anyon model, as treated in this work, all the $R$-matrices can be organized into a single matrix $(R_{a b})_{a,b\in\mathcal{C}}$, which we also refer to as $R$-matrix by slightly abusing the terminology.  

The two different ways of fusing three anyons $a,b$ and $c$ are related by the so called $F$-matrices:
\begin{equation}
\tikzset{every picture/.style={line width=0.75pt}}        
\begin{tikzpicture}[x=0.75pt,y=0.75pt,yscale=-1,xscale=1]
\draw    (87.38,81.17) -- (124.68,118.47) ;
\draw    (161.98,81.17) -- (124.68,118.47) ;
\draw    (124.28,81.57) -- (106.03,99.82) ;
\draw    (124.68,118.47) -- (124.68,156.26) ;
\draw    (276.91,80.53) -- (314.21,117.83) ;
\draw    (351.5,80.53) -- (314.21,117.83) ;
\draw    (313.81,80.93) -- (332.86,99.98) ;
\draw    (314.21,117.83) -- (314.21,155.62) ;
\draw (214.31,123.94) node [scale=0.9] [align=left] {$\displaystyle =\sum _{i\ \left( u,u^{\prime }\right)}\left( F^{d}_{abc}\right)^{i\ \left( uu^{\prime }\right)}_{j\ \left( vv^{\prime }\right)}$};
\draw (125.64,74.01) node  [align=left] {b};
\draw (161.98,74.01) node  [align=left] {c};
\draw (87.38,74.01) node  [align=left] {a};
\draw (314.53,73.37) node  [align=left] {b};
\draw (350.87,73.37) node  [align=left] {c};
\draw (276.27,73.37) node  [align=left] {a};
\draw (320.26,118.15) node [scale=0.7] [align=left] {$\displaystyle u^{\prime}$};
\draw (350,118.15) node  [align=left] {.};
\draw (336.84,100.94) node [scale=0.7] [align=left] {$\displaystyle u$};
\draw (125,163) node  [align=left] {d};
\draw (314.53,162.36) node  [align=left] {d};
\draw (329.83,111.99) node  [align=left] {$\displaystyle i$};
\draw (118.62,120.7) node [scale=0.7] [align=left] {$\displaystyle v^{\prime }$};
\draw (102.05,102.85) node [scale=0.7] [align=left] {$\displaystyle v$};
\draw (109.7,112.63) node  [align=left] {$\displaystyle j$};
\end{tikzpicture}
\end{equation}

In an Abelian anyon model, we have $d=a\otimes b\otimes c$ and the matrix $F^d_{a b c}$ is also reduced to a number $F_{a b c}$. In the toric code model, it is easy to see that $F_{a b c}\equiv1$.  However, this is no longer true for general anyon models.  We measure this value in the paper using `scattering' circuit with one additional ancilla control qubit to show that $F$-matrices are measurable in principle.  And the measurement of nontrivial $F$-matrices will be investigated in future work.

The above mentioned finite set $\mathcal{C}$, fusion rules $N_{ab}^{c}$, $R$- and $F$-matrices uniquely determine an anyon model.


In (2+1)d TQFT, anyons can have fractional spin and statistics. Rotating an anyon $a$ by $2\pi$ (also called twisting) leads to a factor $\theta_a$.
\begin{equation}
\tikzset{every picture/.style={line width=0.75pt}}      
\begin{tikzpicture}[x=0.75pt,y=0.75pt,yscale=-1,xscale=1]
\draw    (164.64,117.61) -- (164.64,140.57) ;
\draw    (164.64,89.6) -- (164.64,110.35) ;
\draw  [draw opacity=0] (166.68,111.6) .. controls (167.66,110.87) and (168.94,110.63) .. (170.13,111.07) .. controls (172.05,111.77) and (173,113.96) .. (172.27,115.97) .. controls (171.53,117.98) and (169.38,119.04) .. (167.47,118.34) .. controls (167.4,118.31) and (167.33,118.28) .. (167.27,118.26) -- (168.8,114.7) -- cycle ; \draw   (166.68,111.6) .. controls (167.66,110.87) and (168.94,110.63) .. (170.13,111.07) .. controls (172.05,111.77) and (173,113.96) .. (172.27,115.97) .. controls (171.53,117.98) and (169.38,119.04) .. (167.47,118.34) .. controls (167.4,118.31) and (167.33,118.28) .. (167.27,118.26) ;
\draw  [draw opacity=0] (164.65,117.95) .. controls (164.63,117.74) and (164.63,117.52) .. (164.63,117.3) .. controls (164.63,114.61) and (165.5,112.33) .. (166.68,111.6) -- (167.5,117.3) -- cycle ; \draw   (164.65,117.95) .. controls (164.63,117.74) and (164.63,117.52) .. (164.63,117.3) .. controls (164.63,114.61) and (165.5,112.33) .. (166.68,111.6) ;
\draw   (167,103.67) -- (164.5,96.67) -- (162,103.67) ; 
\draw    (221.64,117.61) -- (221.64,140.57) ;
\draw    (221.64,89.6) -- (221.64,117.61) ;
\draw   (224,115.67) -- (221.5,108.67) -- (219,115.67) ;
\draw (198,114) node  [align=left] {= $\displaystyle \theta _{a}$};
\draw (158,127) node  [align=left] {a};
\draw (230,126) node  [align=left] {a};
\draw (235,114) node  [align=left] {,};
\end{tikzpicture}
\end{equation}
$\theta_a$ is called topological spin, relating to the ordinary angular momentum spin $s_a$ by $\theta_a = e^{i2\pi s_a}$ and can be determined by $R$- and $F$-matrices through the equation

\begin{equation}
\tikzset{every picture/.style={line width=0.75pt}}
\begin{tikzpicture}[x=0.75pt,y=0.75pt,yscale=-1,xscale=1]
\draw  [draw opacity=0] (230.99,56.16) .. controls (231.94,57.22) and (232.63,58.51) .. (232.94,59.96) .. controls (233.93,64.57) and (230.74,69.07) .. (225.81,70.01) .. controls (220.89,70.95) and (216.1,67.98) .. (215.12,63.37) .. controls (214.52,60.57) and (215.45,57.82) .. (217.38,55.87) -- (224.03,61.66) -- cycle ; \draw   (230.99,56.16) .. controls (231.94,57.22) and (232.63,58.51) .. (232.94,59.96) .. controls (233.93,64.57) and (230.74,69.07) .. (225.81,70.01) .. controls (220.89,70.95) and (216.1,67.98) .. (215.12,63.37) .. controls (214.52,60.57) and (215.45,57.82) .. (217.38,55.87) ;
\draw   (227.82,54.95) .. controls (227.61,53.7) and (227.17,52.65) .. (226.52,51.82) .. controls (227.4,52.44) and (228.5,52.85) .. (229.83,53.05) ;
\draw  [draw opacity=0] (216.88,42.33) .. controls (215.92,41.27) and (215.23,39.97) .. (214.92,38.52) .. controls (213.94,33.91) and (217.13,29.4) .. (222.05,28.46) .. controls (226.97,27.52) and (231.76,30.5) .. (232.75,35.11) .. controls (233.36,37.99) and (232.35,40.82) .. (230.31,42.77) -- (223.83,36.82) -- cycle ; \draw   (216.88,42.33) .. controls (215.92,41.27) and (215.23,39.97) .. (214.92,38.52) .. controls (213.94,33.91) and (217.13,29.4) .. (222.05,28.46) .. controls (226.97,27.52) and (231.76,30.5) .. (232.75,35.11) .. controls (233.36,37.99) and (232.35,40.82) .. (230.31,42.77) ;
\draw    (231.11,56.3) -- (216.75,42.18) ;
\draw    (217.23,56.03) -- (222.08,51.03) ;
\draw    (226.18,47.8) -- (230.47,42.62) ;
\draw (246,53) node [scale=1] [align=left] {$\displaystyle \theta _{a} =\frac{1}{d_{a}} \ \ \ \ \ \ \ \ =\frac{R^{1}_{aa^{*}}}{d_{a} \ \left( F^{a}_{aa^{*} a}\right)^{1}_{1}}.$};
\draw (235.1,51.96) node [scale=0.8] [align=left] {a};
\end{tikzpicture}
\end{equation}

Indeed, the right-hand side of the equation involves a little bit of information about $F$-matrices in a subtle way. In an Abelian anyon model with trivial $F$-matrices, the equation is reduced to a rather simple one
\begin{equation}
\theta_a=R_{aa^*}.
\end{equation}
The effect of twisting (topological spin) is encoded in the modular $T$-matrix through the relation 
\begin{equation}
T_{ab}=\theta_a \delta_{ab}.
\end{equation}

The fractional spins of anyons can be interpreted as their ribbon structure. Expressing anyons with ribbons implies the relation between twisting and braiding $(R^k)^c_{a b} = e^{k\pi i s_a} e^{k\pi i s_b} e^{-k\pi i s_c}I$ where $I$ is the identity matrix of rank $N^c_{a b}$. In the case of $k = 2$, it gives the effect of double-braiding (moving $a$ around $b$ or moving $b$ around $a$ in a full-circle), where $(R^2)^c_{a b} = R^c_{b a} R^c_{a b} = e^{2\pi i s_a} e^{2\pi i s_b} e^{-2\pi i s_c} I = \frac{\theta_a\theta_b}{\theta_c}I$. The data of double-braiding is encoded in the modular $S$ matrix
\begin{equation}
\tikzset{every picture/.style={line width=0.75pt}}   
\begin{tikzpicture}[x=0.75pt,y=0.75pt,yscale=-1,xscale=1]
\draw  [draw opacity=0] (185.94,133.17) .. controls (183.46,135.44) and (180.35,137.07) .. (176.82,137.76) .. controls (166.51,139.77) and (156.51,132.96) .. (154.48,122.56) .. controls (152.46,112.16) and (159.17,102.1) .. (169.48,100.09) .. controls (179.79,98.08) and (189.79,104.89) .. (191.82,115.29) .. controls (192.73,119.99) and (191.86,124.62) .. (189.68,128.47) -- (173.15,118.92) -- cycle ; \draw   (185.94,133.17) .. controls (183.46,135.44) and (180.35,137.07) .. (176.82,137.76) .. controls (166.51,139.77) and (156.51,132.96) .. (154.48,122.56) .. controls (152.46,112.16) and (159.17,102.1) .. (169.48,100.09) .. controls (179.79,98.08) and (189.79,104.89) .. (191.82,115.29) .. controls (192.73,119.99) and (191.86,124.62) .. (189.68,128.47) ;
\draw  [draw opacity=0] (190.02,104.35) .. controls (191.67,102.92) and (193.59,101.76) .. (195.75,100.95) .. controls (205.59,97.27) and (216.57,102.34) .. (220.28,112.26) .. controls (223.99,122.19) and (219.02,133.22) .. (209.19,136.9) .. controls (199.35,140.57) and (188.37,135.51) .. (184.65,125.58) .. controls (182.72,120.4) and (183.14,114.92) .. (185.4,110.34) -- (202.47,118.92) -- cycle ; \draw   (190.02,104.35) .. controls (191.67,102.92) and (193.59,101.76) .. (195.75,100.95) .. controls (205.59,97.27) and (216.57,102.34) .. (220.28,112.26) .. controls (223.99,122.19) and (219.02,133.22) .. (209.19,136.9) .. controls (199.35,140.57) and (188.37,135.51) .. (184.65,125.58) .. controls (182.72,120.4) and (183.14,114.92) .. (185.4,110.34) ;
\draw   (171.28,135.52) .. controls (173.84,136.75) and (176.27,137.32) .. (178.59,137.23) .. controls (176.39,137.97) and (174.32,139.36) .. (172.36,141.41) ;
\draw   (202.36,141.6) .. controls (200.55,139.42) and (198.58,137.88) .. (196.43,136.99) .. controls (198.74,137.24) and (201.21,136.84) .. (203.84,135.8) ;
\draw (223,125) node [scale=1] [align=left] {$\displaystyle S_{ab} =\frac{1}{\mathcal{D}}$ \ \ \ \ \ \ \ \ \ \ \ \ \ \ \ \ \ \ \ \ \
$\displaystyle =\frac{1}{\mathcal{D}}{\displaystyle \sum _{c} N^{c}_{ab}\frac{\theta _{a} \theta _{b}}{\theta _{c}} d_{c},}$};
\draw (162,140) node [scale=0.8] [align=left] {{\footnotesize a}};
\draw (220,139) node [scale=0.7] [align=left] {b};
\end{tikzpicture}
\end{equation}
where $\mathcal{D}$ is the total quantum dimension defined by $\mathcal{D} = \sqrt{\sum_a d_a^2}$. The $S$-matrix is determined by $R$-matrices through the relation
\begin{equation}
S_{a b} = \frac1\CD \sum_c Tr (R^c_{b a}R^c_{a b}) d_c.
\end{equation}

The fusion and braiding are related by the Verlinde formula
\begin{equation}
N_{ab}^{c} = \sum_d \frac{S_{ad} S_{bd} S_{cd}^{*}}{S_{1d}}.
\end{equation}

The above-mentioned relations (6)(9) indicate that the modular matrices $S$- and $T$-matrices can be deduced from $R$- and $F$-matrices.  Since $R$- and $F$-matrices together can uniquely
determine the topological order, we call them the unique identifier of topological order and provide a protocol to measure them.


\subsection{Boundary-bulk duality and anyon condensations}

We demonstrate this protocol by means of quantum simulation using a toric code model with gapped boundaries. As the simplest example of $Z_2$ topological order, the toric code model is a useful platform for demonstrating anyonic statistics~\cite{Kitaev}. 
It is defined on a 2d square lattice consisting of two kinds of plaquettes with qubits on their edges, as illustrated in Fig.~\ref{fig1}c.

The Hamiltonian is a sum of all four qubit interaction terms for the plaquettes (stabilizer operators) in the lattice:
\begin{equation}
H = - \sum_{white\ plaquttes} A_p - \sum_{blue\ plaquettes}  B_p
\end{equation}
The operators $A_p=\Pi_{j\in \partial p} \sigma^x_j$ and $B_p=\Pi_{j\in \partial p} \sigma^z_j$  are
the plaquette operators acting on the four qubits surrounding  white plaquettes and blue plaquettes respectively, as shown in Fig.~\ref{fig1}c $\alpha_1$ and $\alpha_2$. Here $\sigma^x_j$ and $\sigma^z_j$ are the $x$-component and $z$-component of Pauli matrices, respectively,  acting on the $j$-th of the four qubits for each plaquette. All of the above plaquette operators commute with each other, and their eigenvalues are $\pm 1$. 
The ground state (or vacuum, denoted by ${1}$) of the system is the state in which all of the plaquette operators are in $+1$ eigenstates. Anyons can be created in pairs via string operators. As shown in Fig.~\ref{fig1}c $\alpha_3$ and $\alpha_4$, the blue (red) string operator, consisting a sequence of $\sigma^x_j$ ($\sigma^z_j$) operators acting on all qubits in the string, creates a pair of  $m$ anyons ($e$ anyons) on the blue (white) plaquettes at the two ends of the string. 
The fusion of anyons can produce new types of anyons. The fusion rules are expressed as follows: $e \otimes e = m \otimes m = 1$, $e \otimes m = \varepsilon$.

\begin{table*}
	\centering
	\caption{The bulk anyons defined by boundary excitations and the bulk anyon braidings defined by the half braidings of boundary excitations }\label{tab:tab1}
	
	\begin{tabular}{c}
		\\[-1.8ex]
		\\[-1.8ex]
		\hline
		\hline
		\\[-1.8ex]
		
		a) The bulk anyons defined by the boundary excitations and half braidings\\
		\\[-1.8ex]
		\hline
		$1=(1,1\otimes 1=1\overset{1}{\to}1=1\otimes 1,1\otimes e= e\overset{1}{\to}e=e\otimes 1) $ \\
		\hline
		$m= (1,1\otimes 1=1\overset{1}{\to}1=1\otimes 1,1\otimes e =e\overset{-1}{\to}e=1\otimes e) $ \\
		\hline
		$e= (e,e\otimes 1=e\overset{1}{\to}e=1\otimes e, e\otimes e=1\overset{1}{\to}1=e\otimes e)$ \\
		\hline
		$\varepsilon= (e,e\otimes 1=e\overset{1}{\to}e=1\otimes e,e\otimes e=1\overset{-1}{\to}1=e\otimes e) $ \\
		\hline

		\\[-1.8ex]
		\\[-1.8ex]
		\\[-1.8ex]
		\hline
		\hline
		\\[-1.8ex]
		b) The bulk anyon braidings defined by the half braidings of boundary excitations\\
		\\[-1.8ex]
		
		\hline
		$1 \otimes x = x \xrightarrow{c_{1, x}=1} x = x\otimes 1$,  for $x=1, e, m, \varepsilon $ \\
		\hline
		$x\otimes x = 1 \xrightarrow{c_{x, x}=1} 1 = x\otimes x$, for $ x=1, e, m$ \\
		\hline
		$\varepsilon\otimes \varepsilon = 1 \xrightarrow{c_{\varepsilon, \varepsilon}= -1} 1 = \varepsilon\otimes \varepsilon$\\ 
		\hline
		$e\otimes m = \varepsilon \xrightarrow{c_{e,m}=1} \varepsilon = m\otimes e$ \\
		\hline
		$m\otimes e = \varepsilon \xrightarrow{c_{m,e}=-1} \varepsilon = e\otimes m$ \\
		\hline
		$e \otimes \varepsilon =m \xrightarrow{c_{e,\varepsilon}=1} m=\varepsilon\otimes e$ \\
		\hline
		$\varepsilon \otimes e =m \xrightarrow{c_{\varepsilon,e}=-1} m=\varepsilon\otimes e$ \\ 
		\hline
		$m\otimes \varepsilon =e \xrightarrow{c_{m,\varepsilon}=-1} e= \varepsilon\otimes m$ \\
		\hline
		$ \varepsilon\otimes m =e \xrightarrow{c_{\varepsilon,m}=1} e= m\otimes \varepsilon$\\
		\hline
		\hline
		
	\end{tabular}
	\label{talbe1}
\end{table*}

Anyons can be braided. As illustrated in Fig.~\ref{fig1}c $\alpha_5$, when an $m$ anyon is moved around an $e$  anyon along a full circle (double braiding),
 it produces an overall phase of $-1$ between the final and the initial states. This phase relationship  can be encoded by the $S$-matrices as illustrated in Fig.~\ref{fig1}c 
 $\alpha_5$. Another important type of data is the topological spins of the anyons,  encoded in $T$-matrices. The $S$-,$T$-matrices were believed by many to characterize a topological order uniquely, and, in the case of toric code, can be measured by experiments~\cite{Kitaev, RSW, Du}.
It was shown recently that $S$-, $T$-matrices are, however, inadequate to uniquely identify topological order~\cite{MS}. 
For such unique identification,  we need the $R$-matrices (braidings) and $F$-matrices.
The $R$-matrices are actually phase factors for the anyons in toric code model since the anyons here are Abelian. 
Two typical   $R$-matrices for the toric code are $R_{me}^{\varepsilon}=-1$ and $R_{em}^{\varepsilon}=+1$. 
The measurement of the $R$-matrices are the main focus of this article.  

\begin{figure*}
	\centering
	\includegraphics[scale=1.1]{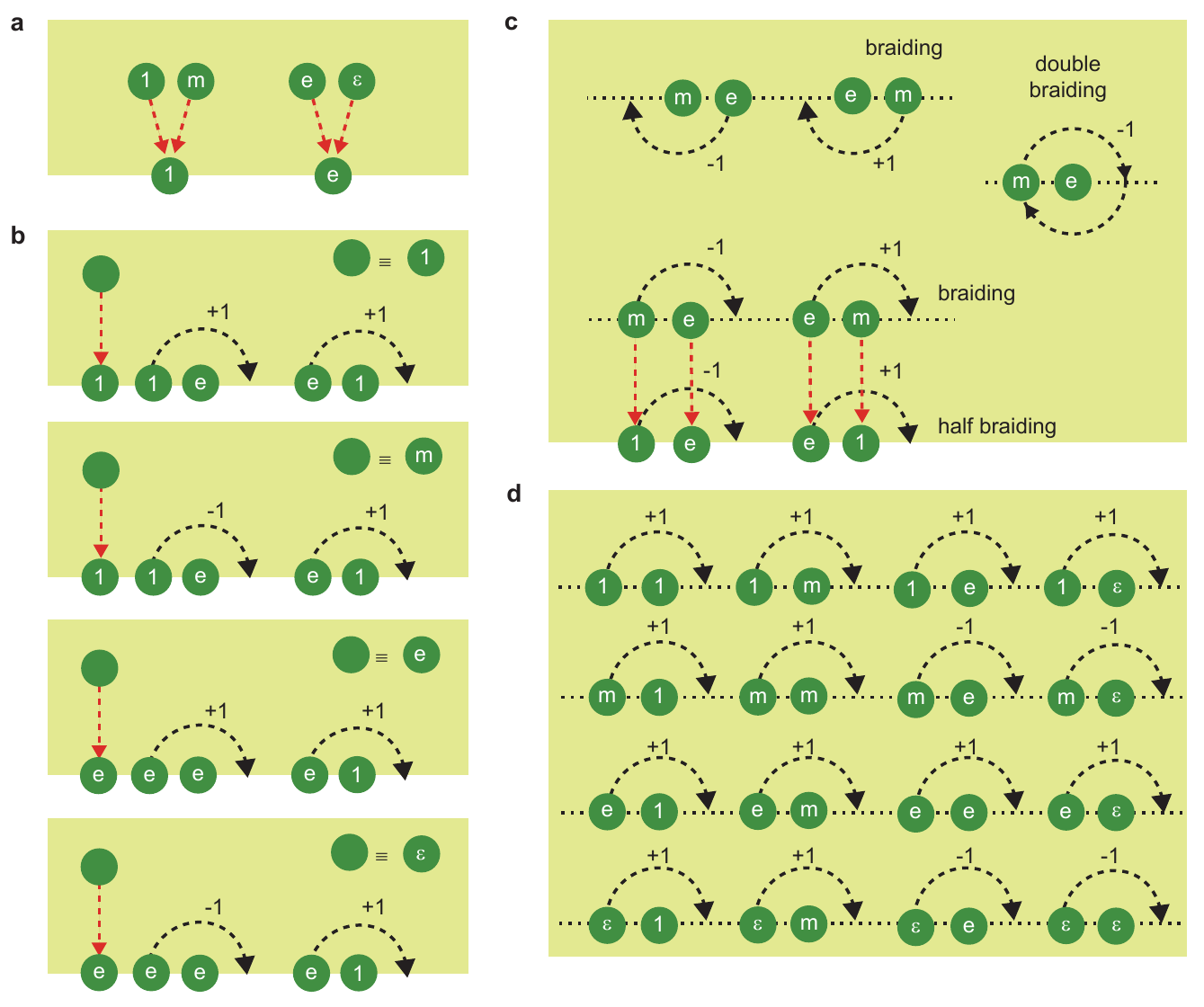}
	\caption{  {\bf Illustrations of half braidings, braidings and double braidings. }  {\bf a}, There are 4 kinds of bulk anyons in the bulk and 2 kinds of  boundary excitations. 
		$m$ and $1$ ($e$ and $\varepsilon$) belong to the same topological sector when they are moved to a white boundary.
		{\bf b}, Definitions of the 4 kinds of bulk anyons using the half braidings on the boundary. $m$ and $1$ are different when they are equipped with
		different half braidings, as are $\varepsilon$ and $e$. 
		{\bf c}, Braidings in the bulk defined in terms of  the half braidings on the boundary.  The black dotted line represents a virtual boundary in the bulk. 
		By moving bulk anyons to the boundary, and using the half braidings of the boundary excitations, the braidings of the bulk anyons can be defined.
		There are two equivalent sets of braidings (the set above the virtual boundary and the set below it) and the double braidings can be obtained from these braidings as well.
		{\bf d},  A complete list of braidings for the 4 kinds of bulk anyons. }
	\label{fig2}
\end{figure*}


The key to measuring the braidings ($R$-matrices) is to make use of the boundaries due to  the boundary-bulk duality~\cite{KK,KWZ} and the anyon condensation on the boundaries~\cite{KL}. There are two topologically distinct boundary types in the toric code~\cite{bk}, namely the white boundaries (known as the smooth boundaries in the original vertex-plaquette form of toric code model) and the blue boundaries (rough boundaries in the original form), as shown in Fig.~\ref{fig1}c. When an $m$ anyon approaches a white boundary, it disappears completely, or equivalently, condenses to the vacuum state. 
So $m$ anyon condense on the boundary. Here, the meaning of ``anyon condensation" is that single anyon can be annihilated or created by local operators.  As we know, local operators doesn't change the topological sectors of the systems. So by local operators, an anyon can only be created with its anit-particle from vacuum. Then to create a single anyon, we need to a series of local operators, i. e. string operators, to seperate it with its anti-particle and push the anti-particle to infinite far way. That is the usual way to create a single anyon by string operators. But due to the existence of boundaries, some kinds of anyons can be created by local operators on the boundaries, which means that they belong to the same topological sector as vacuum. In other words, they condense to vacuum. 
An $e$ anyon cannot pass and becomes a boundary excitation. Therefore, the boundary excitations on a white boundary are $\{ 1,e \}$, and the bulk anyons are mapping to the boundary excitations as $1, m \mapsto 1$, $e, \varepsilon  \mapsto e$ (as illustrated in Fig.~\ref{fig2}a). At a blue boundary,  $e$ anyons disappear and  $m$ anyons remain. Therefore, the blue boundary excitations are $\{1, m\}$ and the corresponding bulk-to-boundary map is $1, e \mapsto 1$, $m, \varepsilon  \mapsto m$.
From above, we can see that bulk-boundary map is not a one-to-one map. There perhaps exists several combinations of boundary excitations for one single bulk topological order. 

It turns out that bulk anyons can be uniquely determined by  the excitations at either of these boundaries~\cite{KK,KWZ,KL}. For example, let us consider a  white boundary, where $m$ anyons condense. 
On this boundary, $m$ and $1$ belong to the same topological sector, i.e. $m=1$. However, when an $m$ is moved into the bulk, it is automatically endowed with additional structures called the {\it half braidings}. These half braidings can be measured by moving the $m$ around a $1$ or $e$ along a semicircle near the boundary, as illustrated by Fig.~\ref{fig1}c $\alpha_7$. It is easy to check that moving an $m$ around a $1$ along a semicircle does not result in a phase difference, whereas moving an $m$ around an $e$ along a semicircle results in a  phase difference of $-1$. Consequence, an $m$ anyon in the bulk can be characterized by the following triple: 
\begin{equation} \label{eq:m}
m= (1,1\otimes 1=1\overset{1}{\to}1=1\otimes 1,1\otimes e =e\overset{-1}{\to}e=1\otimes e),
\end{equation}
where the first component means that the $m=1$ on the boundary and the second and the third components are the half braidings, which are physically measurable quantities. 
Therefore, bulk anyons are actually boundary excitations equipped with half braidings. In this way,
we can recover all four bulk anyons in the forms of four triples (as illustrated in Table.~\ref{talbe1}a and Fig.~\ref{fig2}b).

Furthermore, the braidings among these four anyons can be defined by the half braidings~\cite{egno}. 
For example, we can obtain the following braiding, 
\begin{equation} \label{eq:braiding}
m\otimes e = 1\otimes e \xrightarrow{-1} e\otimes 1 = e \otimes m.
\end{equation}
Here we have used $m=1$ on the boundary and the half braiding $-1$ on the boundary (as illustrated in Fig.~\ref{fig2}c).  
Thus, we obtain $R_{me}^{\varepsilon}=-1$. In addition, since $m$ condenses ($m=1$) on the boundary, $R_{em}^{\varepsilon}=1$ (For $R_{bm}^c$, $b$ is the moving anyon and $m$ is the anyon fixed on the boundary. Since  $m=1$ on the boundary, $R_{bm}^{c}=1$ for an arbitrary $b$ anyon. For  $R_{mb}^{c}$, $b$ is the fixed anyon on the boundary, and $m$ is the moving anyon. Half braiding is applied in this case). 
All of the braidings of bulk anyons can be obtained in this way, as illustrated in Tabel.~\ref{talbe1}b and Fig.~\ref{fig2}c.
Thus, we can obtain the bulk braidings from the boundary half braidings, which are measurable.
Note that this correspondence can be built by considering a virtual boundary in the bulk, and
thus a one-to-one mapping from the bulk braidings to the boundary half braidings can be obtained (as illustrated in Fig.~\ref{fig2}c). 
There are two equivalent sets of bulk braidings, either of which can completely characterize the bulk anyons (as illustrated in Fig.~\ref{fig2}c).
Similarly, we can easily express all braidings among the bulk anyons in terms of  half braidings as illustrated
in Table.~\ref{talbe1}b and Fig.~\ref{fig2}d. Double braidings can be obtained from the bulk braidings by combining the braidings as shown in Fig.~\ref{fig2}c.


In a formal mathematical language, all the above fusion and braiding structures are precisely those of the UMTC $Z(\mathrm{Rep}(\Zb_2))$ and the simple objects in $Z(\mathrm{Rep}(\Zb_2))$ are precisely the four triples given by equations in Table.1. In other words, we obtain the physical proof of the boundary-bulk duality: the bulk excitations, which is given by the UMTC $Z(\mathrm{Rep}(\Zb_2))$, is precisely given by the Drinfeld center of the boundary excitations $\mathrm{Rep}(\Zb_2)=\{ 1, e\}$~\cite{KK,KWZ}. As a byproduct, we have shown that braiding among bulk anyons are physically measurable because the half braidings, which define the braidings in the bulk, are physically measurable! Similarly, we can also use the blue boundary excitations $\{ 1, m\}$ to recover the bulk anyons. Different boundaries $\{ 1, e\}$ and $\{1, m\}$ are Morita equivalent as unitary fusion categories~\cite{KK,KWZ}.  

Toric code mode is a special case of quantum double model (QDM) with group $\mathbb{Z}_2$. Mathematically, the anyons in the quantum double model with group $G$ can be labeled by pairs $(C,\pi)$, the conjugacy classes of $G$ and irreducible representations $\pi$ of the centralizers of $C$~\cite{Cong1, Cong2}. 
It turns out that the excitations on the boundary have a topological order given by a Unitary Fusion Category (UFC). This UFC is the representation category of a quasi-Hopf algebra and is Morita equivalent to the representation category $\mathrm{Rep}(G)$. 
And the elementary excitations in the bulk are simple objects in the UMTC $Z(\mathrm{Rep}(G))$, the Drinfeld center of $\mathrm{Rep}(G)$. It means that the bulk is given by the boundary by taking Drinfeld center~\cite{KL, KWZ}.



\subsection{Measurement scheme of anyon braiding in QDM }
In this section, we propose a measurement scheme of anyon braiding ($R$-matrices) for QDM.
For QDM,  the boundary excitations have a topological order given by UFC $\mathrm{Rep}(G)$ and the  the bulk
anyons are given by UMTC $Z(\mathrm{Rep}(G))$, the Drinfeld center of $\mathrm{Rep}(G)$~\cite{KL, KWZ}. 

When bulk anyons approach the boundaries, some anyons condensed. The condensed anyons form a algebra
$\mathcal{ A} $, so-called the condensation algebra. It is a subcategory of $Z(\mathrm{Rep}(G))$ of which the simple objects
form a commutative algebra~\cite{KL, KWZ}. So anyons in  $a_i\in \mathcal{A} \ (i=1,...,m)$ are bosons and they are commutative, i.e. 
\begin{equation} \label{eq:m}
a_i\otimes a_j= a_j\otimes a_i \ \ (i,j=1,2,...,m).
\end{equation}
In the language of $R$-matrices, $R_{a_i, a_j}=1\ (i,j=1,2,...,m)$.

 After condensation, the survived
anyons on the boundary form a UFC $\mathrm{Rep}(G)$. It is also a commutative algebra no matter group $G$ is
Abelian or non-Abelian. This means that the anyons  $b_k\in \mathrm{Rep}(G)$ $(k=1,2,...n)$ on the boundary are boson and commutative, i.e.
\begin{equation} \label{eq:m}
b_k\otimes b_l= b_l\otimes b_k \ \ \ (k,l=1,2,...,n).
\end{equation}
 In the language of $R$-matrices, $R_{b_k, b_l}=1 \ (k,l=1,2,...,n)$.

 Then arbitrary bulk anyons in UMTC $Z(\mathrm{Rep}(G))$ can be represented by $a_i\otimes b_k\ (i=1,2,...,m; k=1,2,...,n)$.
 We also have $R_{b_k, a_i}=1\ (i=1,2,...,m; k=1,2,...,n)$. The reason is that in this case, anyon $a_i$ are the measured
 anyons which condense to vacuum on the boundary and the circulation of anyon $b_j$ along a semi-circle around a vacuum should not give any
 nontrivial phase factors. While $R_{a_i, b_k}=e^{i\theta_{ik}}\neq 1\ (i=1,2,...,m; k=1,2,...,n)$ (for Abelian anyons) in general since in this case $a_i$ need to move alway from the boundary and they are not vacuum any more. 
 We can see, we only need to measure the $(m-1)(n-1)$ nontrivial phase factors (for Abelian anyons) for QDM to obtain all the $R$-matrices. Other $R$-matrices can be derived from these ones by 
 \begin{equation} \label{eq:m}
(a_i\otimes b_k)\otimes (a_j\otimes b_l)= e^{i\theta_{il}} (a_j\otimes b_l)\otimes (a_i\otimes b_k) 
\end{equation}
 For toric code model, the key nontritival phase factor is $R_{m, e}=-1$.
 
 
We take QDM $D(\Zb_3)$ as an illustrative example for Abelian anyons which also has trivial $F$-matrices. The method of deducing the bulk anyons and boundary excitations for QDM, in general, is discussed in~\cite{Cong2, Anna18}. 
In terms of  charge and flux quantum numbers, the bulk anyons of $D(\Zb_3)$ model are denoted as $\{ 1, e_1, e_2, m_1, m_2, e_1m_1, e_2m_1, e_1m_2, e_2m_2 \}$. The boundary excitations on Type $1$ and Type $2$ boundaries are $\{ 1, e_1, e_2 \}$ and $\{ 1, m_1, m_2 \}$ respectively. 
Two anyons inside the e-class or m-class have trivial mutual statistics, which is indicated by the "Lagrangian subgroup" principle~\cite{Levin13, Xiaoliang13}. The half braidings between anyons in the two different classes (e-class $\{ 1, e_1, e_2 \}$ and m-class $\{ 1, m_1, m_2 \}$) can be obtained through the picture of anyon condensation and boundary-bulk duality similar to the case of toric code discussed above. The four nontrivial phase factors are followings,
 \begin{eqnarray} \label{eq:m}
R_{m1,e_1}=R_{m2,e_2}=\omega \;  \nonumber \\
R_{m1,e_2}=R_{m2,e_1}=\overline{\omega}
\end{eqnarray}
 in which $\omega=\exp(i 2\pi/3)$ and $\overline{\omega}=\exp(-i 2\pi/3)$, other $R$-matrices can be derived from these four factors.
For non-Abelian anyons, the situations are more complicated which is left for future work.

 

The measurement of $ e^{i\theta_{il}} $ 
$(i=1,2,...,m; l=1,2,...,n)$ 
can be obtained by the following procedure:
1) Creating anyon $b_k$ (by string operators) on the boundary as the initial state; 2) Creating anyon $a_i$ on the boundary by local operators;
3) Moving anyon $a_i$ around anyon $b_k$ along a semi-circle and annihilating $a_i$ on the boundary, this state work as the final state; 
4) Comparing the phase difference between the initial state and final state.  

\begin{figure*}
	\centering
	\includegraphics[scale=0.35]{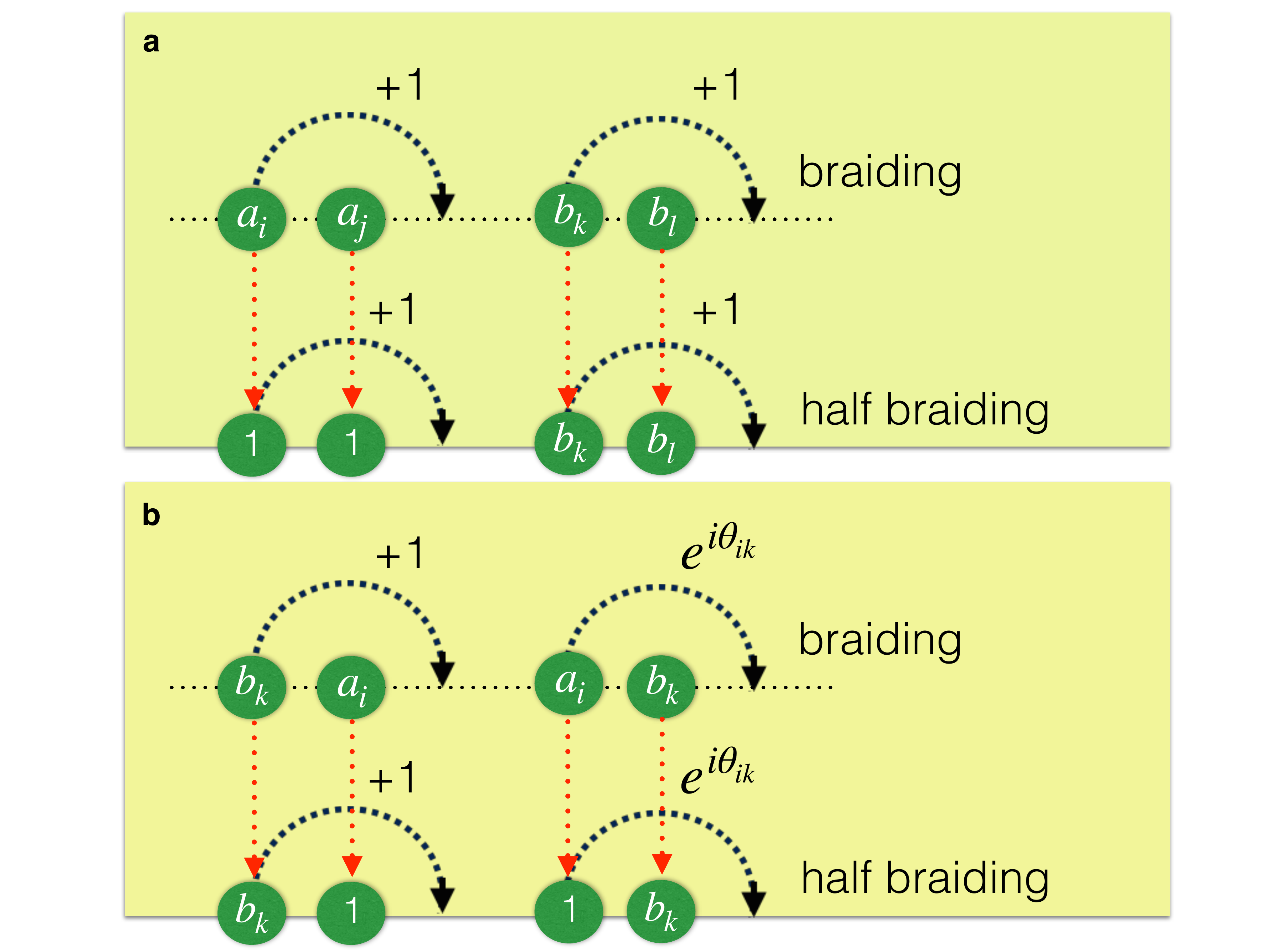}
	\caption{ {\bf Braidings and half braidings in quantum double model. }
		{\bf a}, Anyons in condensed algebra $\mathcal{A}$ are commutative and anyons in UFC $\mathrm{Rep}(G)$ are commutative as well. 
		{\bf b}, $R_{b_k, a_i}=1$ and $R_{a_i, b_k}=e^{i\theta_{ik}}\neq 1$ in general.	}
	\label{fig2b}
\end{figure*}

\subsection{Measurement of the $R$-matrices.}

We use toric code model to demonstrate the measurement of $R$- and $F$-matrices.  
Our experiments are accomplished by means of our NMR quantum computer. We simplify the toric code model in 3, 4-qubit system. 
Our 3, 4-qubit system is a sample of $^{13}$C-labeled trans-crotonic acid molecules dissolved in d6-acetone. The sample consists of four $^{13}$C atoms, as shown in Fig.~\ref{fig5}a, and all experiments are conducted on a Bruker Ascend NMR 600 MHz spectrometer at room temperature.

\begin{figure*}
\centering
\includegraphics[scale=1.2]{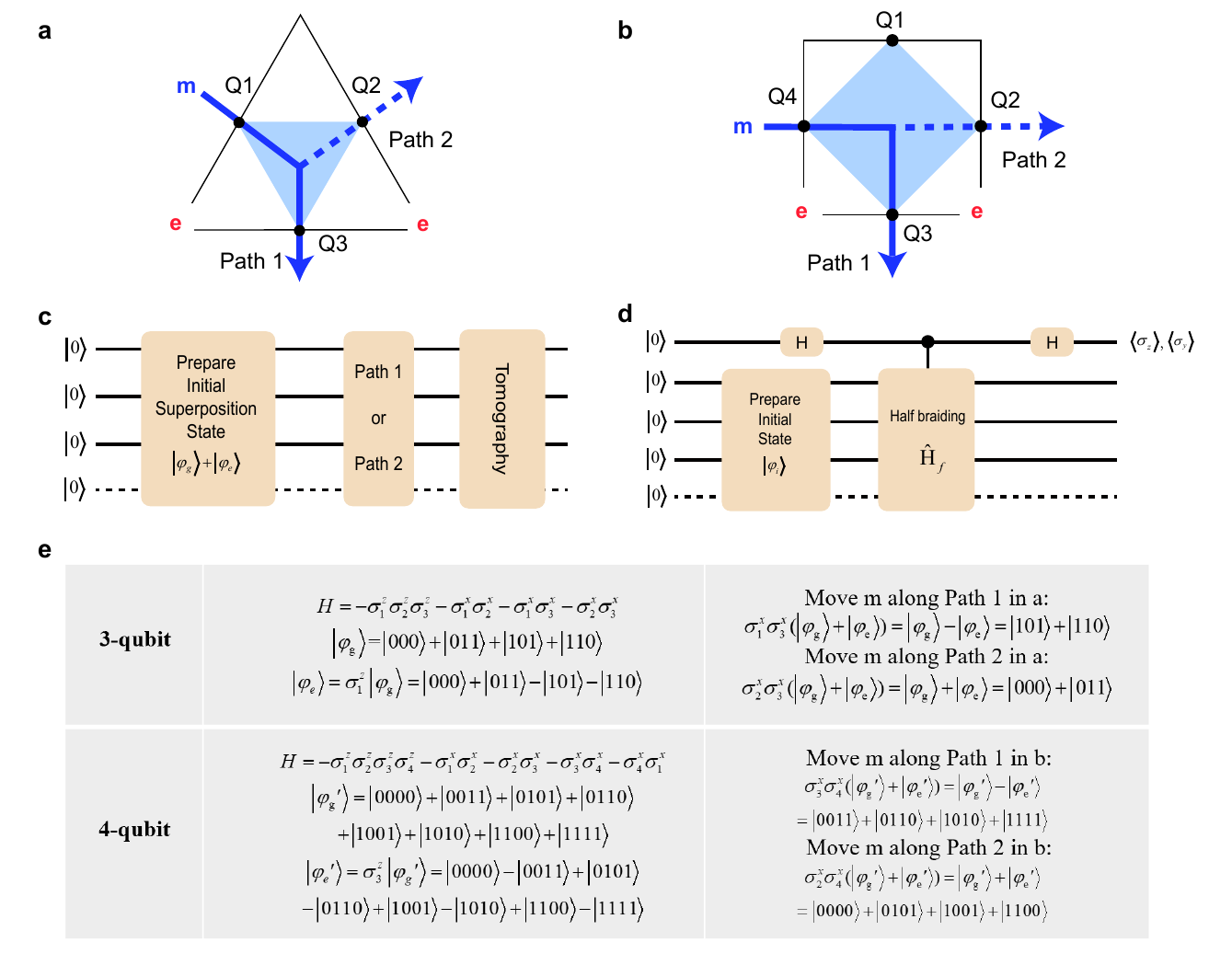}
\caption{ {\bf Illustrations of the experimental setups and corresponding quantum states.}  
  {\bf a}, A 3-qubit toric code model and the half braidings of $m$ along Path $1$ and Path $2$.
{\bf b}, A 4-qubit toric code model and the half braidings of $m$ along Path $1$ and Path $2$.
{\bf c}, Quantum circuit for measuring $m$-$e$ half braidings on  a white (smooth) boundary.
First, the initial state $\left| {{\varphi _{\rm{g}}}} \right\rangle + \left| {{\varphi _{\rm{e}}}} \right\rangle$ is prepared in the first step, where $\left| {{\varphi _{\rm{g}}}} \right\rangle$ is the ground state and $\left| {{\varphi _{\rm{e}}}} \right\rangle$ represents the excited state with two boundary $e$ excitations.
Moving the $m$ anyon through the Path $1$ and Path $2$ by applying a series of $\sigma_x$ operators to the qubits involved in these paths leads to the states $\left| {{\varphi _{\rm{g}}}} \right\rangle - \left| {{\varphi _{\rm{e}}}} \right\rangle$ and $\left| {{\varphi _{\rm{g}}}} \right\rangle + \left| {{\varphi _{\rm{e}}}} \right\rangle$, respectively. These two states can be differentiated via quantum state tomography.
{\bf d}, Quantum circuit for general phase measurement. The state before half braiding is prepared as the initial state $\left| {{\varphi _i}} \right\rangle$, and a half braiding is performed as a controlled operation. For  the general Abelian anyon model, a half braiding operation generates a phase factor, which can be obtained from the two expectation values $\left\langle {{\sigma _z}} \right\rangle$ and $\left\langle {{\sigma _y}} \right\rangle$ on the ancilla qubit. {\bf e}, The Hamiltonians and initial and final states of the half braiding processes  for the 3- and 4-quibits systems. 
}
\label{fig3}
\end{figure*}

In the first experiment, our goal is to show an experimental proof-of-principle demonstration of the half braidings on a gapped boundary and show the effect of the  nontrivial phase factor induced by it. We mainly focus on the measurement of $R_{me}^{\varepsilon}=-1$ since it  is the most important nontrivial phase factor in the toric code. That is, we create a condensed  $m$ anyon at the boundary and move it around a boundary excitation $e$ along a semicircle. The experimental setup and quantum circuits are illustrated in Fig.~\ref{fig3}a-c, and the specific quantum states involved can be seen in Fig.~\ref{fig3}e. The experiment can be divided into three steps: 
1) preparing the initial state, i.e., a superposition of the ground state and the excited state;
2)	performing half braiding by  means of a series of single-qubit rotation operators; and
3)	measuring the final state and using quantum state tomography to obtain the density matrix of the final state.

For the 3-qubit case (Fig.~\ref{fig3}a), the Hamiltonian and the ground state $\left| {{\varphi _{\rm{g}}}} \right\rangle$ of the triangular cell are  shown in Fig.~\ref{fig3}e.
The excited state $\left| {{\varphi _{\rm{e}}}} \right\rangle$ can be obtained via a $\sigma_z$ rotation of qubit $1$ in the ground state leading to two $e$ anyons  on the lower two vertices.  In this system, two different braiding processes can be performed by moving $m$ along either Path $1$ or Path $2$. 
Braiding along Path 1 results in  overall phase factors of  $+1$ for the ground state and $-1$ for the excited state since in the latter case, a half braiding of $m$ around $e$ is performed. In contrast, Path 2 is a trivial path, meaning that braiding along it does not generate any phase factor difference.
To measure the phase factor difference,  we prepare an initial state that is a superposition of the ground state and excited state,
$\left| {{\varphi _{\rm{g}}}} \right\rangle+\left| {{\varphi _{\rm{e}}}} \right\rangle$. Then, half braidings along Path 1 and Path 2 give rise to final states of
$\left| {{\varphi _{\rm{g}}}} \right\rangle- \left| {{\varphi _{\rm{e}}}} \right\rangle$ and $\left| {{\varphi _{\rm{g}}}} \right\rangle +\left| {{\varphi _{\rm{e}}}} \right\rangle$, respectively, which can be identified by quantum state tomography. The 4-qubit case is similar, as shown
in  Fig.~\ref{fig3}b and Fig.~\ref{fig3}e. The corresponding quantum circuits for this experiment are shown in Fig.~\ref{fig3}c.

\begin{figure*}
	\centering
	\includegraphics[scale=1]{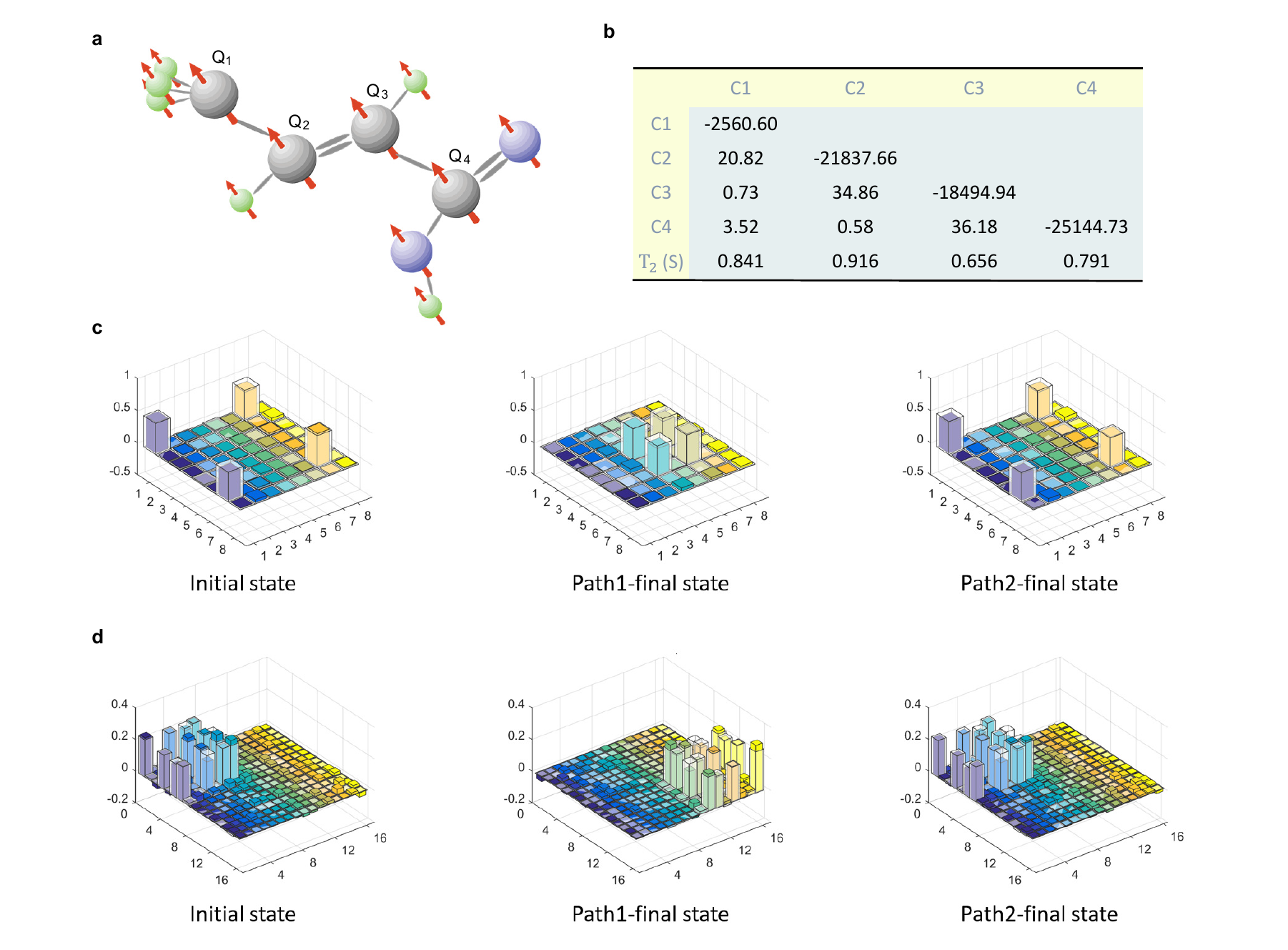}
	\caption{ {\bf The experimental platforms and the results of quantum state tomography. }
		{\bf a}, Our 3, 4-qubit quantum simulator is a sample of $^{13}$C-labeled trans-crotonic acid molecules. We make 4 $^{13}$C atoms from the sample as 4 qubits. 
		{\bf b}, The table on the right lists the parameters of the chemical shifts (diagonal, $\rm Hz$), J-coupling strengths (off-diagonal, $\rm Hz$), and relaxation time scales $T_2$ (seconds).
		{\bf c}, State tomography results for the initial and final states obtained when moving $m$ along different paths in the 3-qubit toric code model. The transparent columns represent the theoretical values, and the colored columns represent the experimental results. Regarding the scale of the X axis in each three-dimensional bar graph, 1 represents the state $|0000\rangle$, 2 represents the state $|0001\rangle$ and so on. Compared with the theoretical results, the two final states in the 3-qubit experiments using Path 1 and Path 2 are obtained with fidelities of 96.37\% and 96.67\%, respectively.
		{\bf d}, State tomography results for the final states obtained by moving $m$ along different paths in the 4-qubit toric code model. Compared with the theoretical results, the two final states in the 4-qubit experiments using Path 1 and Path 2 are obtained with fidelities of 95.23\% and 95.21\%, respectively.
	}
	\label{fig4}
\end{figure*}

In Fig.~\ref{fig4}, we present the state tomography results obtained after running these quantum circuits on our NMR qubit platform.
 We can see that after moving $m$ along Path 2, we obtain a final state that is the same as the initial state.
In contrast, after moving $m$ along Path 1, we obtain a  final state that is completely different from the initial state, 
which is due to the phase factor of $-1$ induced by the half braiding of $m$ around $e$. 
 Thus we obtain the braiding in forms of the $R$-matrices, $R_{me}^{\varepsilon}=-1$.  In principle, all other braidings can be similarly obtained. 
The average fidelities for Path 1 and Path 2 are 96.37$\%$ and 96.67$\%$ for the 3-qubit system, and are 95.23$\%$ and 95.21$\%$ for the 4-qubit system.


In general, a half braiding (denoted as $\hat{H_f}$) leads to a phase factor for Abelian anyons, which can be measured by means of a `scattering' circuit with one additional ancilla control qubit~\cite{Du}, as shown in Fig.~\ref{fig4}d.
The state before half braiding is prepared as the initial state $\left| {{\varphi _i}} \right\rangle$, and the half braiding is performed as a controlled operation. In our experiment, the state before half braiding is prepared as the initial state $\left| {{\varphi _i}} \right\rangle$ with fidelity 95.32\%.
The global phase generated by half braiding is obtained from the two expectation values $\left\langle {{\sigma _z}} \right\rangle$ and $\left\langle {{\sigma _y}} \right\rangle$ on the ancilla qubit through the relations $
\left\langle {{\sigma _z}} \right\rangle  = {\mathop{\rm Re}\nolimits} (\left\langle {{\varphi _i}} \right|\hat{H_f} \left| {{\varphi _i}} \right\rangle )$ and $\left\langle {{\sigma _y}} \right\rangle  = {\mathop{\rm Im}\nolimits} (\left\langle {{\varphi _i}} \right|\hat{H_f} \left| {{\varphi _i}} \right\rangle ) $. 
This method can also be applied to the non-Abelian anyon case, in which the measured values are the matrix elements of the $R$-matrix. 
This circuit is tested for $m$-$e$ half braiding on a 3-qubit plaquette on NMR qubits as a proof-of-principle. $R^{\varepsilon}_{me}=-1$ theoretically corresponds to an exchange statistics phase of $\pi$, and we obtain values of $(1.027\pm 0.001)\pi$ in the experiment.

\begin{figure*}
\centering
\includegraphics[scale=1.1]{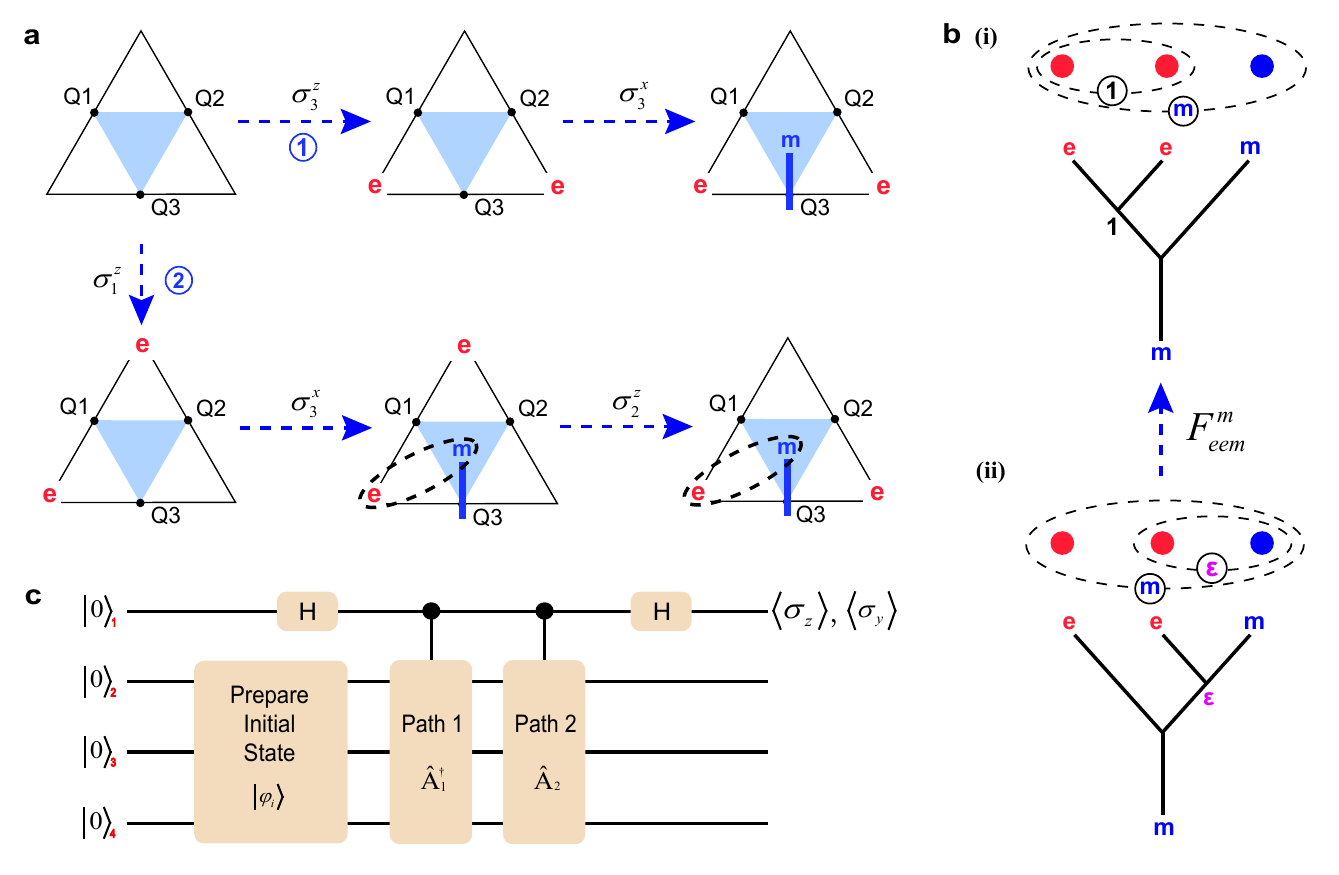}
\caption{ {\bf Two fusion diagrams and the measurement of the $F$-matrix. }
{\bf a}, Two different fusion processes on a 3-qubit plaquette represented by two paths.
{\bf b}, The circle notation and fusion  tree diagrams for Path 1 and Path 2 in (a). 
{\bf c}, The scattering circuit used to measure the overlap of the final states of the two paths, where $\hat{A}_1 = \sigma_3^x \sigma_3^z$,
and $\hat{A}_2 = \sigma_2^z \sigma_3^x \sigma_1^z$. The value of the overlap yields $F_{eem}^{m}$. 
  }
  \label{fig5}
\end{figure*}

\subsection{Measurement of the $F$-matrices.}

The $F$-matrices can be measured by means of a  similar `scattering' circuit~\cite{Du}, as shown in Fig.~\ref{fig5}c.
In this article, we show the measurement of a typical matrix $F_{eem}^m$.

The initial state before fusion is prepared as the ground state $\left| {{\varphi _i}} \right\rangle$ without any anyons.  Two controlled operations $\hat{A}_{1,2}$ are  applied representing two fusion processes using different fusion orders to
fuse two $e$ anyons and an $m$, into an $m$ anyons (illustrated in Fig.~\ref{fig5}a).
The global phase  generated by the different fusion orders, $F_{eem}^m$, is obtained from the two expectation values $\left\langle {{\sigma _z}} \right\rangle$ and $\left\langle {{\sigma _y}} \right\rangle$ of the ancilla qubit. 

In our experiment, we choose $Q3$ in the molecule (Fig.~\ref{fig4}a) as the control qubit to reduce the complexity of the circuit since only  C-not gates between the neighboring qubits are available. $Q1$, $Q2$ and $Q4$ in the molecule represent $Q1$, $Q2$ and $Q3$ in the circuit (Fig.~\ref{fig4}a), respectively. The ground state  $|\psi_g\rangle$ of the 3-qubit toric code model is prepared with a fidelity of $92.96\%$. Then, the operators $\hat{A}_1^\dagger = (\sigma_3^x \sigma_3^z)^\dagger$ and $\hat{A}_2=\sigma_2^z \sigma_3^x \sigma_1^z$ (corresponding to Path $1$ and Path $2$  in Fig.~\ref{fig5}a), which represent  two different orders of fusion of three anyons, are applied to the 3-qubit toric code under the control of qubit $Q4$ . 

We experimentally measure $\left\langle {{\sigma_4 ^z}} \right\rangle$ and $\left\langle {{\sigma_4 ^y}} \right\rangle$  and obtain typical values  of $\left\langle {{\sigma_4 ^z}} \right\rangle = 0.712\pm0.006$ and $\left\langle {{\sigma_4 ^y}} \right\rangle = 0.177\pm 0.004$. We normalize them such that their square sum to $1$ and obtain the angle $\theta=\arctan\left({\left\langle {{\sigma_4 ^y}} \right\rangle}/{\left\langle {{\sigma_4 ^z}} \right\rangle}\right) =(0.077\pm 0.002)\pi$, 
which is close to the theoretical value $0$. 
Thus, we verify that $F_{eem}^{m}=1$ in this experiment.

In the third experiment, to measure  $F$-matrice $F_{eem}^{m}$, we choose $Q3$ in the molecule (Fig.~\ref{fig4}a) as the control qubit to reduce the complexity of the circuit since only  C-not gates between the neighboring qubits are available. $Q1$, $Q2$ and $Q4$ in the molecule represent $Q1$, $Q2$ and $Q3$ in the circuit (Fig.~\ref{fig4}a), respectively. The ground state  $|\psi_g\rangle$ of the 3-qubit toric code model is prepared with a fidelity of $92.96\%$. Then, the operators $\hat{A}_1^\dagger = (\sigma_3^x \sigma_3^z)^\dagger$ and $\hat{A}_2=\sigma_2^z \sigma_3^x \sigma_1^z$ (corresponding to Path $1$ and Path $2$  in Fig.~\ref{fig5}a), which represent  two different orders of fusion of three anyons, are applied to the 3-qubit toric code under the control of qubit $Q4$ . We measure $\left\langle {{\sigma_4 ^z}} \right\rangle$ and $\left\langle {{\sigma_4 ^y}} \right\rangle$ of the control qubit to obtain the overlap of the two final states from two paths,  thus obtain the angle $\theta=\arctan\left({\left\langle {{\sigma_4 ^y}} \right\rangle}/{\left\langle {{\sigma_4 ^z}} \right\rangle}\right)$ and $F_{eem}^{m}=\exp(i\theta)$.


\section{Discussion}
In summary, we experimentally measure  anyon braidings ($R$-matrices) through the boundary-bulk duality,  
that is, bulk anyons are those boundary excitations equipped with half braidings, and  bulk anyon braidings
can be obtained from boundary excitation half braidings. 
Two difficulties that arise in the measurement of $R$-matrices, as noted in the introduction, can be overcome by means of the boundary-bulk duality and the anyon condensation on the boundaries:  
1) if we instead consider the blue boundary, where $e$ anyons condenses, we will obtain another set of the $R$-matrices that is gauge equivalent to what is measured here, and  
2) if an anyon is created on the boundary, half braided with another anyon, and then annihilated on the boundary, the final state and initial state differ only by a phase factor. 
These two difficulties seem to be technical problems, but they are actually  related to the following essential question: what are the fundamental quantities that characterize topological order? This question is similar to the following one: which is the fundamental quantity for an electromagnetic field, the magnetic field/electric field or the vector potential/scaler potential? It would seem that since the vector potential is a gauge-dependent quantity, which means that it is not unique, it should not be measurable. However, the Aharonov-Bohm effect shows that when a particle passes through a region where the magnetic field is zero but the vector potential is nonzero, the phase of the wavefunction is shifted~\cite{AB}. This proves that the vector potential, rather than the magnetic field, is the fundamental physical quantity. The relation between the double braiding and braiding of anyons is similar to that between the magnetic field and the vector potential. For a double braiding of two anyons, the spatial configuration of the final state is the same as that of the initial state, so the effect of the anyonic statistics can be measured by comparing the phase (it is a unitary matrix for non-Abelian anyons)  before and after the double braiding. But for a braiding, i.e. an exchange in positions of two anyons, the spatial configuration of the final state is different from that of the initial state if these two anyons are of different kinds, which makes it unmeasurable in bulk. Furthermore, the $R$-matrices (braidings) are not unique for a given topological order. Therefore, it would seem that double braidings should be the fundamental quantities for topological orders. However, our experiments have demonstrated the important fact that braidings, rather than double braidings, are the fundamental physical quantities for topological orders.

The $F$-matrices can also be  measured using a `scattering' quantum circuit involving the  fusion of three anyons in different orders. 
The $S$-,$T$-matrices can be calculated from the $R$-,$F$-matrices~\cite{MS89,Kitaev06}.
Thus, we provide an experimental protocol for uniquely identifying topological orders.
Although our results are obtained on only a few qubits, the conclusion are applicable to  large systems since the toric code is at a fixed point and the conclusion is independent of the system size~\cite{FP}. 
Furthermore, our boundary-bulk duality between bulk anyons and boundary excitations and the correspondence between bulk anyon braidings and boundary excitation half braidings also holds for other topological orders, even for non-Abelian anyons such as Fibonacci anyons, semions and QDM of $S_3$ group~\cite{KWZ}.
The idea of measuring the half braidings should also be useful to the experimental study of gapless boundaries~\cite{kz19a,kz19b}. 
For a 2d topological order with chiral gapless boundaries, one can apply the folding trick to embed the measurability problem to that of a double-layered system with only gapped boundaries.  such an investigation is left for future work.


\bigskip

\begin{acknowledgments}
	We would like to thank Y.-S. Wu, D. Lu, J. Ye, W. Chen, Z. Tao and Y. Chen for their helpful discussion. J. W. would also like to thank D. Lu for providing the facility of the NMR quantum computing experiments. J. W. was supported by National Natural Science Foundation of China (Grant  No. 11674152 and No.11681240276),  Guangdong Innovative and Entrepreneurial Research Team Program (No. 2016ZT06D348),  Natural Science Foundation of Guangdong Province (Grant No. 2017B030308003) 
	and Science, Technology and Innovation Commission of Shenzhen Municipality (Grant No. JCYJ20170412152620376 and No. KYTDPT20181011104202253).
	J. W., L. K. and H. Z. are supported by the Science, Technology and Innovation Commission of Shenzhen Municipality (Grant No. ZDSYS20170303165926217) and Guangdong Provincial Key Laboratory (Grant No.2019B121203002). L. K. is supported by NSFC under Grant No. 11971219. H. Z. is supported by NSFC under Grant No. 11871078.
	
	
	
\end{acknowledgments}





\bibliography{BraidingPRX_Dec2020}
\bibliographystyle{apsrev4-2prx}

\end{document}